\documentclass[conference,a4paper]{IEEEtran}
\usepackage[utf8]{inputenc}
\usepackage{amsmath,amsfonts}
\usepackage{amssymb}
\usepackage{cite}
\usepackage[final]{graphicx}

\newtheorem{example}{Example}

\newtheorem{theorem}{Theorem}

\newtheorem{definition}{Definition}

\newcommand{\sgn}{\text{sign}}
\def\fF{\mathbb{F}}

\def\cX{\mathcal{X}}
\def\cY{\mathcal{Y}}

\newcommand{\probability}{\ensuremath{\mathbb{P}}}
\newcommand{\Prob}{\probability}

\title{Polarization as a novel architecture to boost the classical mismatched capacity of B-DMCs}
\author{
  \IEEEauthorblockN{Mine Alsan and Emre Telatar}
  \IEEEauthorblockA{EPFL -- I\&C -- LTHI\\
    Lausanne, Switzerland\\
    Email: \{mine.alsan,emre.telatar\}@epfl.ch} 
}
\begin{document}
\maketitle

\begin{abstract}
We show that the mismatched capacity of binary discrete memoryless channels 
can be improved by channel combining and splitting via Ar{\i}kan's polar transformations. We also show that the improvement is possible even if the transformed channels are decoded with a mismatched polar decoder.\\
\end{abstract}


\section{Introduction}\label{sec:intro}

In various communication scenarios, we encounter sub-optimal decoders due to practical implementation constraints (complexity, feasibility requirements) or partial/missing channel information. 
In general, these decoders might perform worse than optimal decoders which minimize the average decoding error probability and possibly result in capacity loss. 
Modeling such sub-optimal scenarios via `reliable communication with a given decision rule' and establishing coding theorems for
them allows one to assess the extent of any loss.

To allow their study within a unified framework, decoders can be categorized based on generic definitions of their decision functions. 
Csisz\'{a}r and Narayan \cite{394641} studied and surveyed the performance of those using additive decision rules: 
Given a codeword set $\left\lbrace \bold{x}(1), \dots, \bold{x}(M)\right\rbrace \in \mathcal{X}^{n}$, an (additive) $d$-decoder assigns a received sequence $\bold{y}$, 
the message $i = 1, \dots, M$ such that $d(\bold{x}(i), \bold{y}) < d(\bold{x}(j), \bold{y})$ for $\forall j\neq i$.  (If there is
no such $i$ the decoder declares an erasure). 
The decision function is computed using the additive extension of a single letter metric $d(x, y)$. Note that the optimal maximum likelihood (ML) decoding rule is included in this class. In addition, a family of decoders of practical interest within the class correspond to mismatch decoders. In this case, the usual ML rule is kept, but instead of the true channel law a different one is employed in 
the decision procedure. 

The transmission capacity of the channel $W$ when decoded with an additive metric $d$ is denoted by $C_d(W)$.  When the metric $d$
corresponds to the ML decoder with respect a channel $V$, we denote the corresponding capacity by $C(W,V)$.  No closed form single letter
expression is known for $C(W,V)$ or $C_d(W)$.  Single-letter lower bounds have been derived, but no converse for any of the lower bounds exists, except for some special cases. Binary input binary output channels are such a case  where $C_{d}(W) = C(W)$ or $0$ 
depending on whether or not the mismatch metric is in `harmony' with the channel behavior \cite{394641}. 
Another exception is the class of binary input discrete memoryless channels (B-DMC). Balakirsky \cite{Balakirsky} gave a computable expression for $C_d(W)$ when $W$ is a B-DMC.
See \cite{394641}, for a more detailed survey on the topic and a more complete list of references.\\

One of the recent breakthroughs in coding theory is Ar{\i}kan's invention of polar codes \cite{1669570}. A polar code of blocklength $N$ is specified by selecting an information set $\mathcal{A}$ over which information is to be transmitted and by freezing the remaining inputs to known values. 
More specifically, one starts by polarizing the channel by the recursive application of the basic polarization transformations defined by Ar{\i}kan. 
The repeated application yields at stage $n$, a set of $N = 2^n$ synthetic channels $\bigl\{W_{N}^{(i)}: i=1, \ldots, N\bigr\}$.
Then, the information set is constructed by selecting the channel indices which are good for uncoded data transmission. Once the data $u_1^N$ is transmitted, the SCD of polar codes will decode the outputs $y_1^N$ using the following estimators \cite{1669570}
\begin{equation}
\hat{u}_{i} =  \left\{\begin{array}{ll}
                                u_{i}, & \hbox{if}  \hspace{3mm} i\in\mathcal{A}^{c} \\
				 f^{(i)}(y_{1}^{N}, \hat{u}_{1}^{i-1}), &\hbox{if}  \hspace{3mm} i\in\mathcal{A}
                                \end{array} \right.,
\end{equation} 
where Ar{\i}kan uses for $f^{(i)}(y_{1}^{N}, \hat{u}_{1}^{i-1})$ the ML decoding rule over the $i$-th synthetic channel $W_N^{(i)}$.

In fact, SCDs can be considered as another large family based on successive cancellation decoding procedures. 
Offering a quite different decoding paradigm than additive decoders, a given SCD will decode the received output $\bold{y}$ in $N$ stages 
using a chain of estimators from $i = 1, \dots, N$ each possibly depending on the previous ones. The estimators $\hat{u}_{i}$ can base their decisions 
on arbitrary single letter metrics of the form $d_i(u_i, y_{1}^{N}\hat{u}_{1}^{i-1})$. 
The SCD of polar codes, however, owes its fame not only for yielding polar coding theorems proving the `symmetric capacity achievingness' of polar codes for a large class of channels, 
but also for inheriting the low complexity structure of the recursive code construction process.

In \cite{MineITW13} and \cite{alsan2013universal}, the performance of polar codes over an unknown B-DMC with mismatched successive cancellation decoding is investigated. The mismatched polar decoder operates according to decision functions $f^{(i)}(y_{1}^{N}, \hat{u}_{1}^{i-1})$ corresponding to the ML rule of mismatched channels $V_N^{(i)}$ synthesized by polarizing a B-DMC $V$ different than the true communication channel $W$. A lower bound denoted by $I(W, V)$\footnote{See Eq. \eqref{eq:I_W_V} for the mathematical definition} is obtained in \cite{alsan2013universal} on the achievable rates by mismatched polar codes over any B-DMC; $I(W, V)$ turns out to be, as the symmetric capacity of the channel \cite{1669570}, a conserved quantity under the polar transforms. While preserving the low complexity structure of the original polar decoder, the mismatched polar decoder extends the theory of channel polarization and polar codes to mismatched processes. 

Let $C_{P}(W, V)$ denote the mismatched capacity of a polar code transmitted over the channel $W$ when decoded with a mismatched polar SCD designed with respect to another channel $V$. The lower bound $I(W, V)$ derived in \cite{alsan2013universal} for $C_{P}(W, V)$ of B-DMCs is also a lower bound to $C(W, V)$, see Fischer \cite{Fischer}. On the other hand, no general order between the polar mismatched capacity $C_{P}(W, V)$ and the classical mismatched capacity $C(W, V)$ can be formulated: there are examples for which $C_P(W,V) > C(W,V)$ and also examples where the reverse inequality holds. 

The main goal of this paper is to show that:
\begin{enumerate}
\item[($g1$)] There are pairs of B-DMCs $W$ and $V$ for which 
$$C(W^+, V^+) + C(W^-, V^-) \geq 2C(W, V)$$ holds.
\item[($g2$)] Furthermore, there exist cases for which $$C_{P}(W, V) > C(W, V)$$ holds.
\end{enumerate} 
For that purpose, we shall study the evolution of the mismatched capacity of B-DMCs under the one-step polarization transformations
when the communication channel and the mismatched channel used in the decision procedure are both symmetrized by the same permutation,
i.e., when for some permutation $\pi$ on the output alphabet satisfying $\pi=\pi^{-1}$, we simultaneously have $W(y|0)=W(\pi(y)|1)$
and $V(y|0)=V(\pi(y)|1)$ for all output letters $y$. \\

The rest of this paper is organized as follows. Section \ref{sec:pre} will briefly introduce the necessary definitions and existing results. Then, some analytical computations and numerical experiments will be presented in the Section \ref{sec:res}. Finally, the paper will close with some discussions. \\

\section{Preliminaries}\label{sec:pre}
We start by presenting the results of \cite{Balakirsky} on the classical mismatched capacity of B-DMCs. 
\begin{definition}\cite{394641}
The $d$-capacity of a discrete memoryless channel $W$, denoted by $C_{d}(W)$, is defined as the supremum of rates $R$ for which, for every $\epsilon > 0$ and sufficiently large block length $n$, 
there exist codes with $d$-decoding such that the code rate $\displaystyle\frac{\log M}{n} > R$, and the maximum probability of error is less than $\epsilon$. 
\end{definition}

Let $W:\mathcal{X}\to\mathcal{Y}$ be a B-DMC with $\mathcal{X} = \{0, 1\}$. We fix an input distribution $P(x)$ on $\mathcal{X}$  and denote the transition probabilities of the channel by $W(y|x)$. Some standard definitions follows:
\begin{equation}
H(PW) = \displaystyle-\sum_{y\in\mathcal{Y}} PW(y)\log _2{PW(y)},
\end{equation}
with $PW(y) = \displaystyle\sum_{x\in\mathcal{X}} P(x)W(y|x)$.
\begin{equation}
H(W|P) = \displaystyle-\sum_{y\in\mathcal{Y}} P(x)W(y|x)\log_2{W(y|x)}.
\end{equation}
\begin{equation}
I(P, W) = H(PW) - H(W|P).
\end{equation}

The following result due to Balakirsky gives a closed form expression for $C_d(W)$ when $W$ is a B-DMC.
\begin{theorem}\cite{Balakirsky}
For any B-DMC $W$ and any additive decoding metric $d(x, y)$
\begin{equation}
C_{d}(W) = \max_{P} I_{d}(P, W),
\end{equation}
where 
 \begin{equation} \label{eq:I_d}
 I_{d}(P, W) \triangleq \displaystyle\min_{\begin{subarray}{l}
		      W': \hspace{2mm} PW'(y) = PW(y) \\
		      \quad \quad d(P, W') \leq  d(P, W)
                    \end{subarray}}
I(P, W'),
\end{equation}
with $d(P, W) = \displaystyle\sum_{x, y} P(x)W(y|x)d(x, y)$.
\end{theorem}

Using this closed from expression, Balakirsky studies the computation of $C_d(W)$ for symmetric B-DMCs when the $d$-decoder preserves the symmetry structure of the communication channel. In the following two examples, we revisit his examples.

\begin{example}\cite[Examples, Statement 1]{Balakirsky}\label{ex:ex1}
For a binary input binary output channel $W$ and any given $d$-decoding rule $I_{d}(P, W) = I(P, W) \mathbf{1}_{\{A\}}$ holds with 
\begin{multline}
A =\{\sgn(1 - W(0|0) + W(1|1)) \\
= \sgn(d(0, 0) + d(1, 1) - d(0, 1) - d(1, 0))\}. 
\end{multline}
So, $C_{d}(W) = C(W)$ or $0$.
\end{example}

\begin{example}\cite[Examples, Statement 2]{Balakirsky}\label{ex:ex2}
Let $W: \mathcal{X}\to\mathcal{Y}$ be a B-DMC with $\mathcal{X} = \{0, 1\}$ and $\mathcal{Y} = \{0, 1, \ldots, L-1\}$. 
Suppose the transition probability matrix of the channel $W$ and the corresponding metrics for the additive $d$-decoder are given by
\begin{equation}
W = \begin{bmatrix}
       w_0 & w_1 & \dots & w_{L-1} \\[0.3em]
       w_{L-1} & w_{L-2} & \ldots & w_0 \\[0.3em]
     \end{bmatrix},
\end{equation}
and 
\begin{equation}\label{eq::d_dec}
d = \begin{bmatrix}
       d_0 & d_1 & \dots & d_{L-1} \\[0.3em]
       d_{L-1} & d_{L-2} & \ldots & d_0 \\[0.3em]
     \end{bmatrix}.
\end{equation}
Then, the mismatched capacity is achieved for $P$ uniform on $\{0,1\}$ and is given by
\begin{equation}
C_d(W) = H(PW) - H(W'|P),
\end{equation}
where $W'$ is given by
\begin{equation}
W' = \begin{bmatrix}
       w_0' & w_1' & \dots & w_{L-1}' \\[0.3em]
       w_{L-1}' & w_{L-2}' & \ldots & w_0' \\[0.3em]
     \end{bmatrix},
\end{equation}
with 
\begin{equation}\label{eq::Cd_channel}
w_y' = \left(w_y+w_{L-1-y}\right) \displaystyle\frac{e^{-\alpha . d_y}}{e^{-\alpha . d_y}+e^{-\alpha . d_{L-1-y}}}, 
\end{equation}
for $y\in\mathcal{Y}$, and the parameter $\alpha\geq 0$ is chosen from the condition:
\begin{equation}\label{eq::alpha_cond}
\displaystyle\sum_{y\in\mathcal{Y}} w_y' d_y = \displaystyle\sum_{y\in\mathcal{Y}} w_y d_y.
\end{equation}
\end{example}

For the rest of this paper, we will restrict the additive decoders to mismatched decoders. Let $V:\mathcal{X}\to \mathcal{Y}$ be a B-DMC symmetrized by the same permutation as the channel $W$ defined in Example 2.
Suppose the transition probability matrix of $V$ is given by
\begin{equation}
V = \begin{bmatrix}
       v_0 & v_1 & \dots & v_{L-1} \\[0.3em]
       v_{L-1} & v_{L-2} & \ldots & v_0 \\[0.3em]
     \end{bmatrix}.
\end{equation}
Then, the corresponding additive decoder can be defined as in \eqref{eq::d_dec} by letting $d_y = -\log{v_y}$, for $y = 0, \ldots, L-1$. 
In this case, the mismatched capacity equals
\begin{equation}
C(W, V) = I(P, W')
\end{equation}
where the transition probabilities of $W'$ can be computed by replacing the relations in \eqref{eq::Cd_channel} and \eqref{eq::alpha_cond} by
\begin{equation}
w_j' = \left(w_j+w_{L-1-j}\right) \displaystyle\frac{\alpha . v_j}{\alpha . v_j + \alpha . v_{L-1-j}}, 
\end{equation}
and
\begin{equation}
\displaystyle\sum_{y\in\mathcal{Y}} w_y' \log{v_j} = \displaystyle\sum_{y\in\mathcal{Y}} w_y \log{v_j}.
\end{equation}
\hspace{2mm}\\

Next, we give the definition of the channel polarization process. 
From two independent copies of a given binary input channel $W:\cX\to\cY$, 
the polar transformations synthesize two new binary input channels $W^-:\cX\to\cY^2$ and
$W^+:\cX\to\cY^2\times\fF_2$ with the transition probabilities given by \cite{1669570}
\begin{align*}
W^-(y_1y_2|u_1)&=\sum_{u_2\in\fF_2}\tfrac12 W(y_1|u_1\oplus u_2)W(y_2|u_2)\\
W^+(y_1y_2u_1|u_2)&=\tfrac12 W(y_1|u_1\oplus u_2)W(y_2|u_2).
\end{align*}
In analyzing the properties of this channel it is useful to introduce \cite{5205856}
an auxiliary stochastic process, $W_0,W_1,\dots,$ defined by
$W_0:=W$, and for $n\geq 0$
$$
W_{n+1}:=\begin{cases}W_n^-&\text{with probability 1/2}\\
W_n^+&\text{with probability 1/2}\end{cases}
$$
with the successive choices taken independently.  In this way, $W_n$ is
uniformly distributed over the set of $2^n$ channels above. \\

Finally, we introduce the results of \cite{alsan2013universal} on the achievable rates by mismatched polar codes. Given two B-DMCs $W$ and $V$, $I(W, V)$ is defined as
\begin{equation}\label{eq:I_W_V}
I(W, V) = \displaystyle\sum_{y}\sum_{x\in\{0,1\}}\frac{1}{2}W(y|x) \log_2{\frac{V(y|x)}{\frac{1}{2}V(y|0)+\frac{1}{2}V(y|1)}}.
\end{equation}
Then, we define the process $I(W_n, V_n)$ associated to the polarization process of
both channels. Note that $I(W_n) \triangleq I(W_n, W_n)$ corresponds to the symmetric capacity process of the channel. The following result shows that $C_P(W, V) \geq I(W, V)$.
\begin{theorem}\label{thm::achievability}\cite{alsan2013universal}
Let $W$ and $V$ be two B-DMCs such that $I(W, V) > -\infty$. Then, $I_\infty(W, V)$ is $\{0, 1\}$ valued with
\begin{equation}\label{eq::pos_rate}
\Prob[I_\infty(W, V)= 1] \geq I(W, V).
\end{equation}
Moreover, the speed of convergence is $O(2^{-\sqrt{N}})$.\\
\end{theorem}

\section{Results}\label{sec:res}
To begin with, we discuss a trivial improvement of the lower bound stated in Theorem \ref{thm::achievability}.
Letting $x^* = \max\{x, 0\}$, the bound in \eqref{eq::pos_rate} can be improved initially as 
\begin{equation}
\Prob[I_\infty(W, V)= 1] \geq I(W, V)^*. 
\end{equation} 
Going a further step, we improve the bound to
\begin{equation}
\Prob[I_\infty(W, V)= 1] \geq \displaystyle\frac{1}{2} I(W^-, V^-)^* + \displaystyle\frac{1}{2} I(W^+, V^+)^*,
\end{equation}
and more generally to
\begin{equation}
\Prob[I_\infty(W, V)= 1] \geq \displaystyle\frac{1}{2^n} \displaystyle\sum_{s\in\{+, -\}^n}I(W^s, V^s)^*,
\end{equation}
for any $n = 0, 1, \ldots$.\\

Now, we study the following example.
\begin{figure}
\centering
\scalebox{1}{\input{mismatch-classical-fig}}
\caption{Classical mismatched decoding.}
\hspace{2mm}\\
\hspace{2mm}\\
\hspace{2mm}\\
\centering
\scalebox{1}{\input{mismatch-pol-fig}}
\caption{One-step polarization architecture for mismatched decoding.}
\hspace{2mm}\\
\hspace{2mm}\\
\hspace{2mm}\\
\centering
\scalebox{1}{\input{mismatch-pol-codes-fig}}
\caption{Polar mismatched decoding.}
\end{figure}
\begin{example}
 Let $W$ be a BSC of crossover probability $\epsilon\in[0, 0.5]$ and $V$ be the BSC of crossover probability $1-\epsilon$. 
  In this example, we will answer the following questions:
\begin{enumerate}
 \item[($q1$)] Suppose we transmit over the channel $W$, but do mismatch decoding with respect to $V$ as shown in Fig. 1. What is the mismatched capacity $C(W, V)$? What is the value of $I(W, V)^*$?
 \item[($q2$)] Suppose we first apply the polar transforms to synthesize the channels $W^+$, $W^-$ and $V^+$, $V^-$, and then we communicate using the architecture given in Fig. 2.
What are the mismatched capacities $C(W^+, V^+)$ and $C(W^-, V^-)$ in this case? What are the values of $I(W^+, V^+)^*$ and $I(W^-, V^-)^*$? 
 \item[($q3$)] Suppose we communicate over the channel $W$ using polar coding, and we do mismatched polar decoding with respect to the channel $V$. The communication architecture is shown in Fig. 3. What is the mismatched capacity of polar coding $C_{P}(W, V)$?\\
\end{enumerate}
Once the answers are derived, we will discuss how this example helps us achieve the two goals we set in the introduction.
\begin{enumerate}
\item[($a1$)] In this case, the crossover probabilities of the BSCs are not in harmony. By the result given in Example \ref{ex:ex1}, we conlcude $C(W, V) = 0$. As $C(W, V) \geq I(W, V)$, we have $I(W, V)^* = 0$.
 
\item[($a2$)] It is known that after applying the minus polar transform to a BSC of crossover probability $\alpha\in[0, 1]$, 
the synthesized channel is also a BSC, and with crossover probability $2\alpha(1-\alpha)$. So, both $W^-$ and $V^-$ are the same BSC with crossover probability of $\epsilon^- = 2\epsilon(1-\epsilon)$.
Therefore, the mismatched capacity of the minus channel equals its matched capacity, i.e., $C(W^-, V^-) = C(W^-) = 1-h(\epsilon^-)$. Similarly, $I(W^-, V^-)^* = C(W^-) = 1-h(\epsilon^-)$.

\item[($a3$)] We have already seen that $V^-=W^-$.  It is easy to see that while $V^+\neq W^+$, one has $V^{+-}=W^{+-}$, and indeed,
$V^{++-}=W^{++-}$, \dots.  Consequently, for any sequence $s^n=s_1,\dots,s_n$ of polar transforms $V^{s^n}=W^{s^n}$
except when $s^n=+\dots+$.  Thus, $I(W^{s^n},V^{s^n})=I(W^{s^n})$ for all $s^n$ except $+\dots+$, and we see
that $C_P(W,V)=C(W)$, and we arrive at our second goal (g2).\\
\end{enumerate}
\end{example}

After such a motivating example, one is curious about whether the improvement we illustrated for the specific pair of BSCs extend to other pairs of mismatched B-DMCs as well. To satisfy one's curiosity, we now present the results of the numerical experiments we carried for random pairs of channels with various output alphabet sizes.
Let 
\begin{equation}
\Delta(W, V) \triangleq C(W^+,V^+)+C(W^-,V^-)-2C(W,V).
\end{equation}
The numerical experiments show that:
\begin{itemize}
\item[(i)]
When the output alphabet is binary or ternary, one-step improvement, i.e. $\Delta(W, V)>0$, happens only when $C(W,V)=0$ and can be as large as $1/2$.  When $C(W,V)>0$, we observe that $\Delta(W, V) = 0$, and one has neither improvement nor loss. See Fig. 
4.

\begin{figure}
\centering
\scalebox{0.92}{\input{delta-bala3-support-fig}}
\caption{$C(W, V)$ versus $\Delta(W, V)$ for $|\mathcal{Y}| = \{2, 3\}.$}
\hspace{2mm}\\
\scalebox{0.92}{\input{delta-bala4-support-fig}}
\caption{$C(W, V)$ versus $\Delta(W, V)$ for $|\mathcal{Y}| = 4.$}
\end{figure}

\item[(ii)]
When the output alphabet contains four or more symbols, improvement may happen not only when $C(W,V)=0$ but also when $C(W,V)>0$;
however, there are cases when $C(W^+,V^+)+C(W^-,V^-)<2C(W,V)$, so one may encounter a loss, i.e. $\Delta(W, V)<0$, after a one-step transformation. See Fig. 5.
\end{itemize}
Finally, the numerical experiments also suggest that 
\begin{equation}
\displaystyle\sum_{s\in\{+,-\}^n}\frac{1}{2^n} I(W^s, V^s)^* \leq C(W, V)
\end{equation}
holds whenever $C(W, V) > 0$, (we did experiment only for $n= 1, 2, 3$). Thus, 
$C_P(W, V) \leq C(W, V)$ is a likely conjecture for the case where $C(W, V) > 0$.  \\
 
\section{Discussions}\label{sec:dis}
The study by Balakirsky \cite{Balakirsky} gave the initial impulse for this work: We adapted his example \cite[Examples, Statement 2]{Balakirsky} which computes $C_d(W)$ of a symmetric B-DMC $W$ when the additive decoder shares the symmetry structure of $W$ to mismatched decoders, and we carried numerical experiments based on this computation to compare $C(W, V)$ with various other quantities such as the sum $C(W^+, V^+) + C(W^-, V^-)$. The experiments reveal that, as opposed to $I(W, V)$, $C(W, V)$ is not necessarily a conserved quantity under the polar transformations. Nevertheless, communication rates higher than $C(W, V )$ can be achieved in some cases by integrating the polarization architecture of Ar{\i}kan into the classical mismatched communication scenarios. Furthermore, by studying a specific pair of BSCs $W$ and $V$ for which $C(W, V) = 0$, but $C_{P}(W, V) > 0$, we showed that there exists channels for which the polar transformations strictly improve the mismatched capacity of B-DMCs. \\

\section*{Acknowledgment}
This work was supported by Swiss National Science Foundation under grant number 200021-125347/1.

\bibliographystyle{IEEEtran}
\bibliography{ref}

\end{document}